\documentclass[showpacs,aip,jap,amsmath,amssymb,reprint]{revtex4-1}
\usepackage{graphicx}
\usepackage{bm}
\usepackage{color}

\begin{document}

\title{Electronic structure, optical and magnetic properties of Co$_{2}$FeGe Heusler alloy films}

\author{N. V. Uvarov}

\author{Y. V. Kudryavtsev}
\affiliation{Institute of Metal Physics, National Academy of
Sciences of Ukraine, Vernadsky 36, 252680, Kiev-142, Ukraine}

\author{A. F. Kravets}
\email{anatolii@kth.se}
\affiliation{Nanostructure Physics, Royal Institute of Technology, 10691 Stockholm, Sweden}
\affiliation{Institute of Magnetism, National Academy of Sciences of Ukraine, Vernadsky 36 b, 03142 Kiev, Ukraine}

\author{A. Ya. Vovk}
\author{R. P. Borges}
\author{M. Godinho}
\affiliation{CFMC, Department of Physics, University of Lisbon, Campo Grande, Edif. C8, Lisbon, 1749-016 Lisbon, Portugal}

\author{V. Korenivski}
\affiliation{Nanostructure Physics, Royal Institute of Technology, 10691 Stockholm, Sweden}

\begin{abstract}

Optical properties of ferromagnetic half-metallic full-Heusler Co$_{2}$FeGe alloy are investigated experimentally and theoretically.
Co$_{2}$FeGe thin films were obtained by DC magnetron sputtering and show the saturation magnetization at $T$=10 K of $m\approx$5.6 $\mu_{B}$/f.u., close to the value predicted by the Slater-Pauling rule. First-principles calculations of the electronic structure and the dielectric tensor are performed using the full-potential linearized-augmented-plane-wave method in the generalized gradient (GGA) and GGA+U approximations. The measured
interband optical conductivity spectrum for the alloy exhibits a
strong absorption band in the 1 - 4 eV energy range with
pronounced fine structure, which agrees well with the calculated
half-metallic spectrum of the system, suggesting a near perfect
spin-polarization in the material.

\end{abstract}

\pacs{71.20.$\pm$b, 75.30.$\pm$m, 78.20.$\pm$e}

\maketitle

\section{Introduction}

Heusler alloys (HA) attract significant attention due to their interesting physical properties promising various practical applications in the fields of smart materials and magneto-electronics.\cite{graf} Indeed, certain Co$_{2}$Fe-based full-Heusler alloys (FHA) are predicted to be ferromagnetic and half-metallic, \emph{i.e.}, exhibiting near 100\% spin polarization of the charge carriers at the Fermi level. This property makes Co$_{2}$Fe-based FHA's (and Co$_{2}$FeGe among them) promising candidates to be used as spin-injectors in spintronic devices. Furthermore, Co$_{2}$Fe-based FHA's have the highest Curie temperatures among the known half-metallic ferromagnets. Therefore, fabrication and investigation of the electronic structure and physical properties of Co$_{2}$Fe-based FHA films are of fundamental and technological interest.

The Co$_{2}$FeGe system and, in particular, films are rather unexplored. Only a few publications discuss the physical properties of non-stoichiometric Co-Fe-Ge alloy films [(CoFe)$_{0.70}$Ge$_{0.30}$, (CoFe)$_{100-x}$Ge$_{x}$] and spin-valves based on them. \cite{cfgfilm1, cfgfilm2,cfgfilm3} Full-Heusler $L$2$_{1}$-type ordered Co$_{45}$Fe$_{31}$Ge$_{24}$ thin films were successfully fabricated by Takamura,\emph{et al.} using a thermally activated germanidation reaction between ultra-thin germanium-on-insulator and Co-Fe layers deposited on it.\cite{experim} However, such films showed magnetic moments of $m$=4.8$\mu_{B}$, which was notably smaller than the bulk value of $m$=5.54$\mu_{B}$\cite{buschow}, probably due to their off-stoichiometry.

The important aspect of the half-metallic Heusler alloys is their unique magneto-optical (MO) properties: a discovery of giant Kerr rotation in half-Heusler PtMnSb alloy opens the way for applications in MO reading-recording.\cite{engen} Optical and MO properties of a number of half-metallic FHA's have been investigated theoretically and experimentally.\cite{kubo,pico1,pico2,shreder1,kumar,shreder2,shreder3,nader,xu,kim} It has been shown that the predictions significantly depend on the calculation methods  employed.\cite{kumar,kim} At the same time, the experimental optical and MO properties of the HA's are often quite difficult to explain without first-principle calculations. It is preferred, therefore, to combine the two approaches for a comprehensive analysis of a specific material, explaining the measured data at the same time as verifying experimentally the theoretical model used. 
Optical and magneto-optical properties of Co$_{2}$FeGe compound
were have been simulated by Kim \emph{et al.} using the generalized
gradient approximation (GGA) and GGA+U approaches.\cite{kim} It
was shown that the predicted properties significantly depend on
calculation approach used, so a comparison with the experiment was necessary to validate and, if necessary, refine the alternatives. Therefore, one of the article's purpose is checking the correctness of various calculation tools
by the comparison of calculated and experimental physical (optical
and magnetic) properties of alloy.

In this work we fabricate the largely unexplored Co$_{2}$FeGe alloy in the film form using sputtering, measure their optical and magnetic properties,
and develop and refine first-principles models for explaining the data observed. Our results indicate half-metallicity, which should be interesting for a number of applications. 

\section{Experimental and theoretical methods}

Co$_{2}$FeGe Heusler alloy films of 50 nm in thickness were
deposited at 500 $^{\circ}$C on thermally oxidized Si substrates
using DC magnetron co-sputtering from Co$_{2}$Fe and Ge targets.
The base pressure in the deposition chamber was 5$\times$10$^{-8}$
Torr and the Ar pressure used during deposition was 5 mTorr. The
deposition rates for Co$_{2}$Fe and Ge components were 0.0781 and
0.0174 nm/sec, respectively. The film composition was
determined using x-ray dispersion spectroscopy analysis.

The crystalline structure was analyzed using X-ray
diffraction (XRD) in the $\theta-$2$\theta$ geometry with
Cu-K$_{\alpha }$ radiation of 1.5406 \AA\ wavelength.

Magnetic measurements were performed using a SQUID MPMS-5 magnetometer in the temperature range of 5-300 K. Magnetic field was applied in the film plane. The substrate contribution was subtracted using the methods described by Garcia \emph{et al.}\cite{garcia}

The electronic structure,  optical properties such as the dielectric function (DF), and and magnetic properties of the Co$_{2}$FeGe alloy with $L2_{1}$ type sructure (space group $Fm\overline{3}m$) were calculated using the WIEN2k
code\cite{blaha}, utilizing a full-potential linearized-augmented-plane-wave method with GGA and GGA+U methods.\cite{wimmer} For the exchange-correlation functional, the generalized-gradient-approximation version of Perdew \emph{et al}.\cite{perdew} was used. In the $L2_{1}$-type structure Co atoms are located at (1/4, 1/4, 1/4), Fe atoms at (0.0, 0.0, 0.0) and Ge atoms at (1/2, 1/2, 1/2) in the unit cell of a 225
space group.

The cell size was determined in the calculations from minimizing the full electron energy versus the lattice parameter. Self-consistency was obtained using 816 k-points in the irreducible Brillouin zone (IBZ). To calculate the DF of Co$_{2}$FeGe we used 4285 k-points in the IBZ.

The calculated optical properties of the Co$_{2}$FeGe HA were
compared with the experimental ones. Optical properties, such as 
[$Re({\sigma}) = \omega\varepsilon_{2}/4\pi$ and
$\varepsilon_{1}$, where $\sigma$ is the optical conductivity
(OC), $\varepsilon_{1}$ and $\varepsilon_{2}$ the real and
imaginary parts of the diagonal components of the DF
$\widetilde{\varepsilon}=\varepsilon_{1}-i\varepsilon_{2}$] of the
samples were measured using a spectroscopic rotating analyzer-ellipsometer, at 293 K, in the spectral range of 310 - 2500 nm (4.0 - 0.5 eV) at a fixed incidence angle of 73$^{\circ}$.

\section{Results and discussion}

\begin{figure}[t,p]
\includegraphics [width=8.5cm]{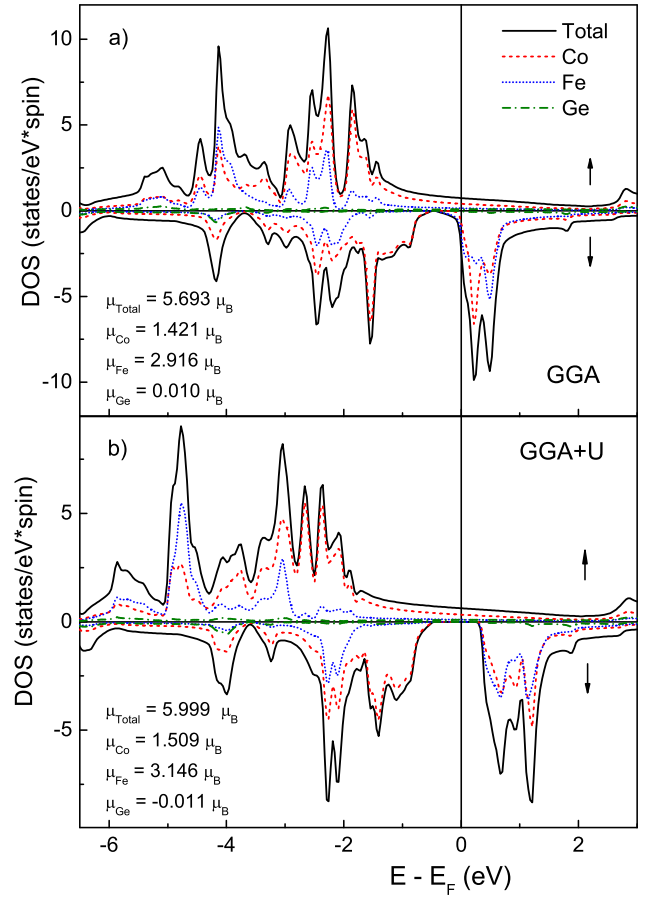}
\caption{Spin-resolved DOS for $L2_{1}$-phase of Co$_{2}$FeGe FHA, calculated using a) GGA  and b) GGA+U approximations.}
\label{fig1}
\end{figure}

The calculated density of the electronic states (DOS) for Co$_{2}$FeGe alloy
with the $L2_{1}$-type crystal structure and the lattice parameter of $a$=5.75 \AA, obtained using the GGA and GGA + U approaches, are shown in Fig. \ref{fig1}. It is seen that the main contributions to the
resulting DOS of Co$_{2}$FeGe alloy obtained by either GGA or GGA+U
methods are due to the Co and Fe atoms. The Co and Fe states are
hybridized - the most intense peaks of the DOS are formed by
the coincident in energy Co and Fe states. A detailed
analysis shows that the contribution to the hybridized states is due mainly to the Co and Fe 3$d$ states, which indicates a covalent
character of their interaction. The contribution to the total DOS from
the Ge states is small. This means that the Ge atoms form essentially
ionic bonds with the surrounding atoms.

For the minority bands both approaches reveal a deep minimum (GGA) or 
even an energy gap of about 1 eV  at the Fermi level (GGA+U). Thus,
GGA+U predicts Co$_{2}$FeGe HA to be a half-metal with the spin polarization of $P$=1.000, while GGA predicts still a rather high $P$=0.504.

The DOS shown in Fig. \ref{fig1} agree well with the results in the literature obtained for $L2_{1}$-type ordered Co$_{2}$MnGe using LDA, LDA+U, GGA, and GGA+U.\cite{kumar,balke,kumar2,kim}

The GGA and GGA+U methods reveal somewhat different values of the
magnetic moment of the alloy. According to the Slater-Pauling
rule,\cite{slater,pauling} L2$_{1}$-type Co$_{2}$FeGe HA should have a magnetic moment of $\mu_{Co2FeGe}$ = $N$ - 24 = 6$\mu_{B}$, where $N$= 2Co (3$d^{7}4s^{2}$)+Fe(3$d^{6}4s^{2}$)+Ge(4$s^{2}4p^{2}$)= 30. The GGA+U result for the magnetic moment is almost exactly that expected from Slater-Pauling,  $m_{Co_{2}FeGe}=5.999 \mu_{B}$, Co=1.509$\mu_{B}$, Fe=3.146$\mu_{B}$, Ge=-0.011$\mu_{B}$, while GGA yields $m_{LDA}=5.693 \mu_{B}$ (see Fig. \ref{fig1} b).

Approximately the same difference between the magnetic moments of
Co$_{2}$FeGe calculated using LDA (GGA) and LDA+U (GGA+U) can be found in the literature. Thus, Kandpal \emph{et al.} report the total magnetic moment of the alloy of $m_{Co_{2}FeGe}$=5.72 $\mu_{B}$/f.u. (Co=1.42$\mu_{B}$,
Fe=2.92$\mu_{B}$).\cite{kandpal} Kumar \emph{et al.} calculate for Co$_{2}$FeGe the total magnetic moment of $m_{Co_{2}FeGe}$=5.72 $\mu_{B}$/f.u.
(Co=1.43$\mu_{B}$, Fe=2.88$\mu_{B}$, Ge=0.01$\mu_{B}$) and 6.02 $\mu_{B}$/f.u.
(Co=1.53$\mu_{B}$, Fe=3.12$\mu_{B}$, Ge=-0.03$\mu_{B}$), for GGA and GGA+U, respectively.\cite{kumar,kumar2} These values for the magnetic moments in Co$_{2}$FeGe, obtained using both LDA and GGA, are in good agreement with the experimental values: $m_{exp}$=5.54, 5.70 or 5.74 $\mu_{B}$/f.u.\cite{buschow,sargo,kumar2}

Taking into account that the magnetic moment of Co$_{2}$FeGe obtained 
using the GGA+U approach fits better the Slater-Pailing rule than does the moment in the GGA approximation, we calculate the optical properties of Co$_{2}$FeGe using GGA+U.

Because of rather small Co$_{2}$FeGe alloy films thickness
(50 nm) their composition was examined by employing both
energy-dispersive X-ray spectrometer of JEOL JSM-6490LV and
wavelength dispersive X-ray spectrometer of JEOL JXA-8200.
Aforementioned tools revealed a composition (in atomic percent) of Co$_{55.7}$Fe$_{24.0}$Ge$_{20.3}$ and Co$_{59.7}$Fe$_{26.5}$Ge$_{13.8}$ (hereafter Co$_{2}$FeGe), respectively.
 
\begin{figure}[t,p]
\includegraphics [width=8.5cm]{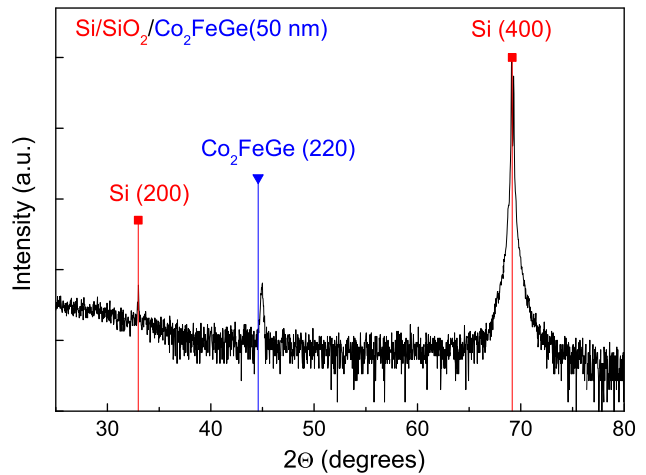}
\caption{XRD pattern of a Co$_{2}$FeGe film deposited on a thermally oxidized Si substrate.}
\label{fig2}
\end{figure}

The X-ray diffraction pattern for a Co$_{2}$FeGe film deposited on a thermally oxidized Si substrate is shown in Fig. \ref{fig2}. Additional to the reflection peaks of the Si substrate, only one diffraction peak of Co$_{2}$FeGe is present. It corresponds to a (220) diffraction peak of a cubic structure and connected with a non-epitaxial growth of the Co$_{2}$FeGe film.\cite{experim} The deposited film grows on the amorphous SiO$_{2}$ surface that does not provide favourable conditions for epitaxy. The lattice parameter derived using Bragg's law is $a$=5.702 \AA. This value is slightly lower than those determined experimentally for bulk stoichiometric Co$_{2}$FeGe alloy
($a$=5.738 \AA or $a$=5.743 \AA).\cite{buschow,balke0} Applying the volume  correction to our data yields $a$=5.75 \AA.

\begin{figure}[t,p]
\includegraphics [width=8.5cm]{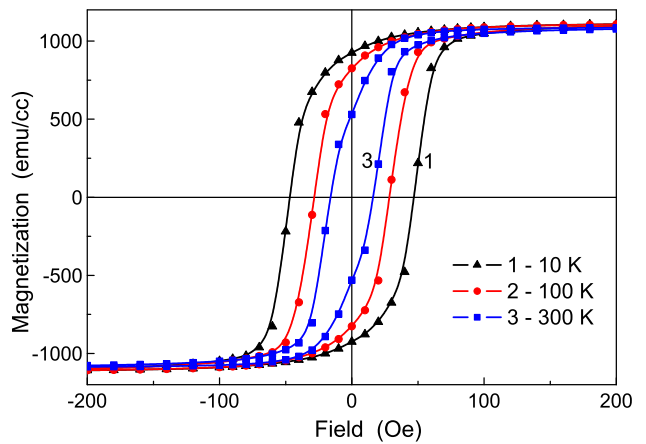}
\caption{Magnetic hysteresis loops of a Co$_{2}$FeGe film measured
at different temperatures with the magnetic field applied in the film
plane.}
\label{fig3}
\end{figure}

Magnetic hysteresis loops for a Co$_{2}$FeGe film  measured at different temperatures are shown in Fig. \ref{fig3}. The film demonstrates ferromagnetic properties in the whole investigated temperature range. The
saturation magnetization practically does not change with temperature from 10 to 300 K, decreasing only by $\sim$3\% (from
1120 emu/cc to 1090 emu/cc). Such behavior of magnetization-vs-temperature indicates a high Curie temperature $T_{C}$ of the alloy. Taking into account that Co$_{2}$FeGe FHA have four atoms
per unit cell and applying the experimental values of the saturation
magnetization ($M_{s}$) at $T$=10 K and the lattice parameter one
obtains $M_{s}\approx$5.6 $\mu_{B}$/f.u.
This value is very close to
that predicted by the Slater-Pauling rule (\emph{i.e.} 6.0
$\mu_{B}$/f.u.) as well as that calculated using the GGA+U approach.
It should also be pointed out that Co$_{2}$FeGe films have
relatively low coercivity, which increases from $H_{c}\sim$15 Oe
to 45 Oe when the temperature decreases from 300 down to 10 K.
This fact implies fine-grain crystalline structure of the samples.

On the basis of the rather good agreement between the
experimental and the calculated magnetic moments of Co$_{2}$FeGe
alloy one can expect that this alloy should exhibit a high degree of spin polarization, predicted by the GGA+U results.

\begin{figure}[t,p]
\includegraphics [width=8.5cm]{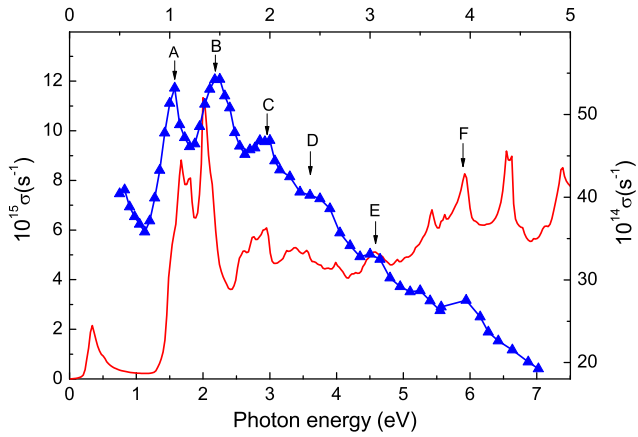}
\caption{Experimental (triangles, right and top scales) and calculated
(line, left and bottom scales) optical conductivity spectra of Co$_{2}$FeGe alloy film.}
\label{fig4}
\end{figure}

\begin{figure}[t,p]
\includegraphics [width=8.5cm]{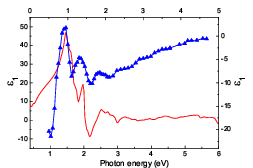}
\caption{Experimental (triangles, right and top scales) and
calculated (line, left and bottom scales) $\varepsilon_{1}$
spectra of Co$_{2}$FeGe alloy film.}
\label{fig5}
\end{figure}

Figures \ref{fig4} and \ref{fig5} show the measured and
calculated optical properties of the Co$_{2}$FeGe alloy. The
experimental optical conductivity spectrum exhibits a strong
absorption peak in the 0.5 $ < \hbar\omega < $ 3.5 eV energy range, with superposed smaller peaks marked by A - F (see Fig. \ref{fig4}). The theoretical  optical properties were obtained using the GGA+U approach. A visual comparison of the experimental $\sigma(\hbar\omega)$ spectrum with calculated
one allows us to conclude that they are in good qualitative agreement. A slight
shift in energy is the common occurrence originating from some
uncertainty of determining the ground level energy in the GGA+U calculation. A good qualitative agreement between the experimental and calculated spectra
is found also for the real part of the diagonal components of the
DF, $\varepsilon_{1}(\hbar\omega)$, shown in Fig. \ref{fig5}. The intraband contributions to the calculated $\sigma(\hbar\omega)$ and $\varepsilon_{1}(\hbar\omega)$ were not included, which can be the cause of some differences in the peak positions between the calculated and experimental
$\sigma(\hbar\omega)$ spectra.

\begin{figure}[t,p]
\includegraphics [width=8cm]{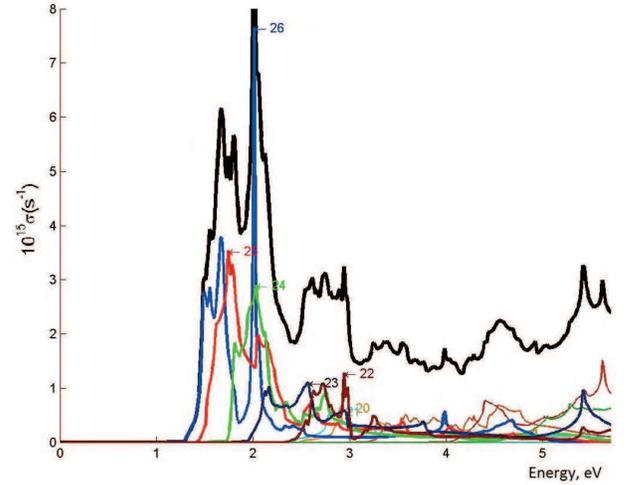}
\caption{Calculated total (thick black solid line) and partial
contributions to the interband optical conductivity spectrum,
produced by electron excitations within the spin-down (minority) sub-bands of the Co$_{2}$FeGe FHA.}
\label{fig6}
\end{figure}

\begin{figure}[t,p]
\includegraphics [width=8cm]{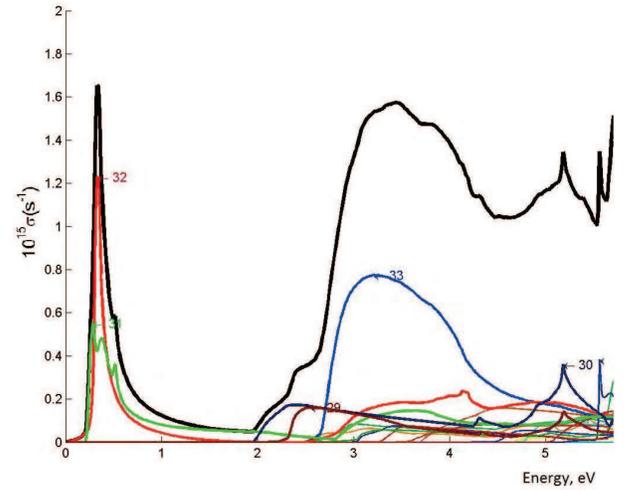}
\caption{Calculated total (thick black solid line) and partial
contributions to the interband optical conductivity spectrum,
produced by electron excitations within the spin-up (majority) sub-bands of the
Co$_{2}$FeGe FHA.}
\label{fig7}
\end{figure}

\begin{figure}[t,p]
\includegraphics [width=8.5cm]{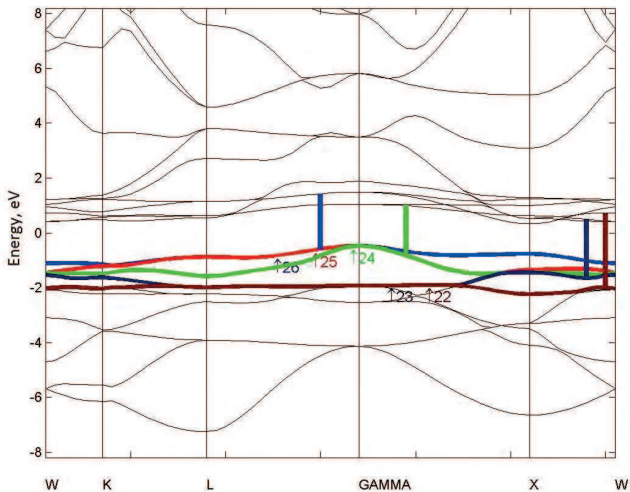}
\caption{Spin-down (minority) band structure of Co$_{2}$FeGe FHA.
The most intense electron transitions are marked by arrows.}
\label{fig8}
\end{figure}

\begin{figure}[t,p]
\includegraphics [width=8.5cm]{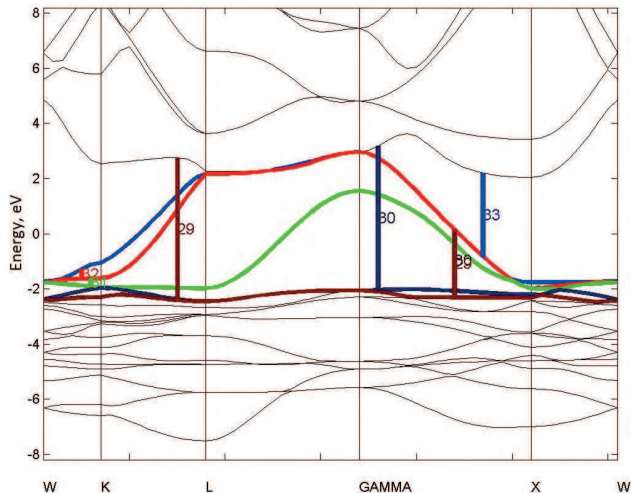}
\caption{Spin-up (majority) band structure of Co$_{2}$FeGe FHA. The most
intense electron transitions are marked by arrows.}
\label{fig9}
\end{figure}

Figures \ref{fig6} and \ref{fig7} show the calculated interband
optical conductivity spectra with partial contributions from the
electron excitations from the various occupied minority and majority
bands to unoccupied ones. The main contributions to the most intense
experimental interband absorption peaks A and B in Fig.
\ref{fig4} are due to the electron transitions from the 24$^{th}$,
25$^{th}$ and 26$^{th}$ minority bands (Fig. \ref{fig6}) to all unoccupied bands. These transitions take place between nearly
parallel bands in the vicinity of high symmetry points L and
$\Gamma$, shown in Fig. \ref{fig8}. Peak C at
$\hbar\omega$=2 eV in the experimental optical conductivity
spectrum (Fig. \ref{fig4}) results mainly from the electron
excitations from the majority 33$^{rd}$ band (Fig. \ref{fig7}) to
all unoccupied bands in the vicinity of the high-symmetry point K (Fig. \ref{fig9}). The interband transitions in the minority bands have
sharp edges at $\hbar\omega$=1.2 eV.

Unlike electron transitions within minority bands, there is rather
intense and very narrow absorption peak at $\hbar\omega$=0.5 eV
(see Fig. \ref{fig7}) due to the electron excitations from the
31$^{st}$ and 32$^{nd}$ bands within majority bands (Fig.
\ref{fig9}). This peak is not observed experimentally because it is
outside of the measuring range. Absorption peaks D - F in the experimental
$\sigma(\hbar\omega)$ spectrum (Fig. \ref{fig4}) are due to electronic transitions in both the minority and majority bands, as can be seen in Figs. \ref{fig6} and \ref{fig7}).

The experimental $\varepsilon_{1}(\hbar\omega)$ spectrum shows generally
the Drude-like behavior, \emph{i.e.}, increase in the absolute value being negative with decreasing photon energy. This general dependence is modulated by the anomalous dispersion regions near the most intense absorption peaks due to electronic transitions in the system (Fig. \ref{fig5}).

\section{Conclusions}
Nearly stoichiometric Co$_{2}$FeGe Heusler alloy films have been fabricated using DC-magnetron co-deposition from Co$_{2}$Fe and Ge targets. The saturation magnetization of the material is measured to be close to that predicted by the Slater-Pauling rule. The measured optical properties are well explained in terms of the alloy's band structure, calculated ab-initio using the GGA+U approach, and found to correspond to a half-metallic ferromagnet. The half-metallicity of the obtained material may prove useful for applications in spin-polarizers and spin-injectors in magnetic nanodevices.

\section*{Acknowledgments}
We gratefully acknowledges support from project EU-FP7-FETOpen-STELE, support by the Ukrainian National program "Fundamental problems of nanostructural systems, nanomaterials and nanotechnologies" project 28/12-H. Work of A.Ya. Vovk and P.R. Borges was supported by Portuguese FCT through "Ciencia 2008" and "Ciencia 2007" programs, respectively.

\end{document}